%% file: EnergyEfficientNetworks.tex
\documentclass[10pt,conference]{IEEEtran}
\usepackage{cite}

\input{macros.tex}

\title{Base station selection for energy efficient network operation with the majorization-minimization algorithm}

\author{
\IEEEauthorblockN{Emmanuel Pollakis\IEEEauthorrefmark{1},
		Renato L. G. Cavalcante \IEEEauthorrefmark{1},
		S\l awomir Sta\'nczak\IEEEauthorrefmark{1}\IEEEauthorrefmark{2}}
		\IEEEauthorblockA{
	\IEEEauthorrefmark{1}
		Fraunhofer Institute for Telecommunications, Heinrich Hertz Institute,
		Einsteinufer 37,  10587 Berlin, Germany,\\
    \IEEEauthorrefmark{2}
    Heinrich-Hertz-Lehrstuhl f\"ur Informationstheorie und theoretische Informationstechnik,\\
    Technische Universit\"at Berlin, Einsteinufer 27, 10587 Berlin, Germany\\		
	 Email: \{emmanuel.pollakis, renato.cavalcante slawomir.stanczak\}@hhi.fraunhofer.de
    }
}

\begin{document}
\maketitle

\begin{abstract}
In this paper, we study the problem of reducing the energy consumption in a mobile 
communication network; we select the smallest set of active base stations 
that can preserve the quality of service 
(the minimum data rate) required by the users. In more detail, we start 
by posing this problem as an integer programming problem, the solution of which shows 
the optimal assignment (in the sense of minimizing the total energy consumption) between base stations and users. In particular, this solution shows which base stations can then be switched off or put in idle mode to save energy. However, solving this 
problem optimally is intractable in general, so in this study we develop a suboptimal 
approach that builds upon recent techniques that have been successfully applied to, 
among other problems, sparse signal reconstruction, portfolio optimization, statistical 
estimation, and error correction. More precisely, we relax the original integer 
programming problem as a minimization problem where the objective function is concave and 
the constraint set is convex. The resulting relaxed problem is still intractable in 
general, but we can apply the majorization-minimization algorithm to find good solutions 
(i.e., solutions attaining low objective value) with a low-complexity algorithm. 
In contrast to state-of-the-art approaches, the proposed algorithm can take into account inter-cell interference, 
is suitable for large-scale problems, and can be applied to heterogeneous networks 
(networks where base station consume different amounts of energy). 

\end{abstract}

\section{Introduction}
\label{sec:intro}  
The information technology sector contributes to an increasingly portion of the 
world's energy consumption, and thus there is an urge to improve the energy efficiency in 
communication networks. By improving the energy efficiency, network operators also reduce 
operational costs because energy constitutes a significant part of their expenditures. 
Recent studies \cite{Willkom2009UserBehavior, CorlHufschmid2008} have shown 
that there are large load fluctuations in time and space in mobile networks; the traffic 
demand is low at night time and high during working hours. Therefore, there is a huge 
potential to save energy by adapting the network to the demanded traffic. Unfortunately, 
current networks are typically configured to provide the best possible quality of service 
(QoS) by assuming that the largest expected traffic is always demanded. This assumption 
often implies that all base stations should be powered at all times, thus wasting too 
much energy because base stations are one of the most energy expensive components of 
a mobile cellular network (they consume over 50 \% of the total energy budget 
\cite{HanHarrold2011}). \par 

Against this background, current work has been considering to minimize 
the number of base stations to provide a given quality of service to users 
 in order to save energy \cite{Brevis2011,Niu2010CellZooming}. In particular, the work 
in \cite{Brevis2011} has proposed a scheme to minimize the energy consumption by 
optimizing the number of base stations and their locations. The problem is posed 
as mixed integer programming problem, and the authors suggest to solve it with the 
simplex method and the branch and bound algorithm. Although this scheme has been 
originally proposed to find a fixed, non-adaptive network configuration, it can be 
easily extended to the case where the network configuration has to be modified 
according to changes in traffic demands during the day. However, this work focuses 
on the time division multiple access (TDMA) protocol, so it does not consider, for 
example, inter-cell interference, one of the major problems in modern systems 
\cite{Majewski2010}.  In addition, algorithms based on branch and bound methods are 
known to run for a very long time, even with problems of moderate sizes \cite{joshi09}.
In contrast, the work in \cite{Niu2010CellZooming} has proposed centralized and 
decentralized algorithms to address specifically the problem of base station selection 
in the present of traffic fluctuations in the network. These algorithms are fast, 
but they are based on heuristics and do not consider networks where base stations 
have different power consumption (which, in particular, is the case of modern networks 
consisting of hierarchical structures). 
Furthermore, no analytical justification is provided to 
support the good performance of the algorithms, and the dynamic power consumption of the 
base stations is also not considered.

To address the limitations of the above techniques, we propose an algorithm that 
tries to select the smallest number of base stations needed to provide a required 
data rate to all users in the system.  In more detail, we model the base station 
selection problem as an integer programming problem, which is known to be intractable 
for large systems (thus we cannot expect to solve this problem optimally). Therefore, 
to find a fast (but not necessarily optimal) solution, we use ideas similar to those 
successfully applied in sparse optimization based on convex programming 
\cite{joshi09,candes08b,yamada11}. In more detail, building upon the results in 
\cite{candes08b,sri11}, we relax the integer programming problem by posing it as 
the minimization of a concave function constrained to a convex set, and we obtain 
a base station configuration attaining low objective value by using the 
majorization-minimization (MM) algorithm \cite{hunter04}. In doing so, we are able
to devise an algorithm that is fast, has an analytical justification for its good 
performance, can easily consider heterogeneous networks, and can take into account 
inter-cell interference and the transmitted power of base stations.

The remainder of this paper is organized as follows. 
In Section~\ref{sec:model} we present the system model. In 
Section~\ref{sec:SparseOptiEnergyEffi} we show the optimization problem and derive 
the proposed algorithm, which is evaluated empirically in 
Section~\ref{sec:numericalResults}. 
\section{Scenario and System Model}
\label{sec:model}
In this study, we consider a representative urban cellular network with a dense
base station deployment. 
In more detail, we denote the set of all base stations as $\setm = \left\{ 1,2,...,M 
\right\}$ and the set of all users in a cellular radio network as $\setn = \left\{ 
1,2,...,N \right\}$.  
As in \cite{Niu2010CellZooming}, the channel state information (CSI) of 
the channels, and hence their spectral efficiency, is available at the base stations. 
We also assume that the spectral efficiency is known for any link at a 
central unit, and there is no intra-cell interference (the latter assumption is common in network planning of modern systems \cite{Majewski2010}). To account for inter-cell interference, we assume
the worst case interference when computing the spectral efficiency. 
More precisely, we approximate the spectral efficiency of the link from base station 
$i$ to user $j$ by 
\begin{empheq}{align}
\omega_{i,j} = \eta^{\text{BW}}_{i,j} \log \left(1+\frac{P_{i,j}}{\eta^{\text{SINR}}_{i,j} 
\left(\sum_{d \neq i}{P_{d,j}}+P_j^{\text{noise}} \right)} \right)
\label{eq:specEffPL}
\end{empheq}
where $P_j^{\text{noise}}$ is the noise power for user $j$ and $\eta^{\text{BW}}_{i,j}$; $\eta^{\text{SINR}}_{i,j}$ are suitable scaling factors (known as the bandwidth efficiency and
signal-to-interference-plus-noise ratio efficiency, respectively) \cite{Majewski2010}; and $P_{i,j}$ is the received signal power from base station $i$ to user $j$, which is determined by the ITU log-distance path loss model with 
shadow fading for urban macro cell environments \cite{3GGPTR36814}.

All base stations report their CSI and the QoS requirements of the users to a 
central unit, which also knows the spatial traffic load distribution.
All users have fixed QoS requirements represented by a minimum required data rate 
$\ve{r} = \left[ r_1,r_2, \ldots, r_N \right]$. 
To support this data rate, base station $i$, to which user $j$ is connected, has 
to allocate bandwidth $b_{i,j} = \frac{r_j}{\omega_{i,j}}$, where 
$\omega_{i,j}$ is the $i$th row and $j$th column entry of the spectral efficiency 
matrix $\mbox{\boldmath$\omega$} \in \R^{M \times N}$ holding the spectral efficiency 
for all links.
In addition to the QoS constraints for the users, each base station $i$ has only a 
limited amount of bandwidth $B_i$ to allocate to its users. 

The problem we study in this paper is to find the set of base stations consuming 
the smallest amount of energy while providing the desired QoS level to each user.
In the next section we formalize this problem and provide an efficient solution.

\section{Sparse Optimization for Increased Energy Efficiency}
\label{sec:SparseOptiEnergyEffi}
Let $\ma{X} \in \left\{ 0,1 \right\}^{M \times N}$ be a 0-1 matrix where
$x_{i,j}$ denotes its $i$th row and $j$th column. The parameter $x_{i,j}$ is $1$ if user 
$j$ is connected to base station $i$ and $0$ otherwise.
To save energy by finding the minimum number of base stations necessary to provide the 
minimum QoS (data rate) to every user, we have to solve the following optimization 
problem:
\begin{empheq}{align}
\underset{\ma{X}\in\R^{M\times N}}{\mbox{min. }} & c || \ma{X} \cdot \ones ||_0 + 
f_{\text{\boldmath$\omega$},\ve{r}}(\ma{X})&  
\label{eq:originalProbEP}\\
\mbox{subject to  } & \sum^N_{j = 1} \frac{r_j}{\omega_{i,j}}x_{i,j} \leq B_i & 
 i \in \setm \label{eq:originalProb1EP}\\
& \sum^M_{i = 1} x_{i,j} = 1 &  j   \in \setn \label{eq:originalProb2EP}\\
& x_{i,j} \in \left\{ 0,1 \right\} &  i \in \setm,  j   \in \setn,  
\label{eq:originalProb3EP}
\end{empheq}
where $|| \cdot ||_0$ denotes the $l_0$-norm (the number of non-zero elements) and 
$\ones \in \R^N$ denotes the vector of ones. 
(We assume that there is at least one feasible solution to problem 
(\ref{eq:originalProbEP})-(\ref{eq:originalProb3EP}).)
The parameter $c$ is the static energy consumption per base station\footnote{Here
we assume that all base stations 
consume the same amount of power when active. 
However, in strong contrast to \cite{Niu2010CellZooming}, our scheme can easily be 
adapted to the case where base stations consume different amounts of power.}, 
and the function $f_{\text{\boldmath$\omega$},\ve{r}}(\ma{X})$ is a suitable concave or convex 
function accounting for the dynamic energy consumption depending on the load at the
base station. 
In the following, for the sake of simplicity, we neglect the dynamic part because, 
with current technology, 
the dynamic energy consumption is marginal compared to the static energy
consumption for a typical base station \cite{CorlHufschmid2008}. (This assumption also 
enables us to compare the proposed scheme with the cell zooming approach in \cite{Niu2010CellZooming}.)
The constraints in (\ref{eq:originalProb1EP}) guarantee that users do not exceed the 
bandwidth available to the base stations, whereas the constraints in 
(\ref{eq:originalProb2EP}) force every user to be connected to exactly one base station.
Each row of $\ma{X}$ corresponds to a base station, so we seek solutions 
$\ma{X}^\star$ with as many rows having all-zero entries as possible. 
These rows correspond to the base stations that will be deactivated.
Note that problem (\ref{eq:originalProbEP})-(\ref{eq:originalProb3EP}) is a combinatorial 
problem, and thus intractable if $M N$ is large. Building upon the results 
in  \cite{candes08b}, we devise a low-complexity
algorithm that tries to solve a problem strongly related to 
(\ref{eq:originalProbEP})-(\ref{eq:originalProb3EP}) and that is able to provide 
matrices $\ma{X}$ with good row sparsity patterns. 
To do so, we first rewrite (\ref{eq:originalProbEP}) in a more convenient form for our 
purposes. 

For mathematical convenience, define $\ve{w}:=\mathrm{vec} 
(\ma{X})=:[w_1~\ldots~w_{MN}]^T\in\R^{MN}$ where $\mathrm{vec}(\cdot)$ is the operator vectorizing
a matrix by stacking its columns.
For any given $\ve{h}=[h_1\ldots h_M]^T\in\R^M$, we can verify the following 
simple relation \cite{candes08b,sri11}:
\begin{align*}
 \|\ve{h}\|_0=\lim_{\epsilon\to 0}\sum_{i=1}^M \dfrac{\log(1+|h_i| \epsilon^{-1})}{
\log(1+\epsilon^{-1})}.
\end{align*}

Therefore, problem (\refeq{eq:originalProbEP})-(\ref{eq:originalProb3EP}) 
(ignoring $f_{\text{\boldmath$\omega$},\ve{r}}(\ma{X})$ for the reasons described above) is 
equivalent\footnote{Here we say that two optimization problems are equivalent
if the set of solutions is the same.} to
\begin{empheq}{align}
 \underset{\ve{w}\in\R^{MN}}{\mbox{min. }} & \lim_{\epsilon\to 0}\sum_{i=1}^M \dfrac{
\log(1+\epsilon^{-1} \ve{s}_i^T\ve{w} )}{\log(1+\epsilon^{-1})} &  \nonumber\\
 \mbox{s. t. } & \sum^N_{j = 1} \frac{r_j}{\omega_{i,j}}w_{i+M(j-1)} \leq B_i & 
  i \in \setm \nonumber \\
 & \sum^M_{i = 1} w_{i+M(j-1)} = 1 &  j   \in \setn  \nonumber\\
 & w_{i} \in \{0,1\}& i\in\{1,\ldots,MN\}, \nonumber
 \end{empheq}
where $\ve{s}_i=\mathrm{vec}(\ma{S}_i)\in\R^{MN}$ and $\ma{S}_i\in\R^{M
\times N}$ is a matrix of zeros, except for its $i$th row, which is a row of ones. 
For mathematical tractability, we relax the above problem by fixing $\epsilon>0$ 
and by using a convex relaxation of the last constraint.\footnote{The relaxation 
also gives rise to adaptions to coordinated multi-point transmission/reception
(CoMP) scenarios, where users are allowed to connect to multiple base stations
simultaneously.}
In doing so, we obtain the following problem:
\begin{empheq}{align}
 \underset{\ve{w}\in\R^{MN}}{\mbox{min. }} & \sum_{i=1}^M \dfrac{\log(\epsilon+ 
\ve{s}_i^T\ve{w} )-\log(\epsilon)}{\log(1+\epsilon^{-1})} &  \label{eq:relax}\\
 \mbox{s. t. } & \sum^N_{j = 1} \frac{r_j}{\omega_{i,j}}w_{i+M(j-1)} \leq B_i & 
  i \in \setm \label{eq:c1} \\
 & \sum^M_{i = 1} w_{i+M(j-1)} = 1 &  j   \in \setn  \label{eq:c2}\\
 & w_{i} \in [0,1] &  i \in \{1,\ldots,MN\}. \label{eq:c3}
 \end{empheq}
Alternatively, problem (\ref{eq:relax})-(\ref{eq:c3}) (ignoring unnecessary constants)
can be expressed more compactly as:
\begin{empheq}{align}
\label{eq:reducedProblem}
 \underset{\ve{w}\in\mathcal{X}}{\mbox{min. }} & \sum_{i=1}^M \log(\epsilon+ 
\ve{s}_i^T \ve{w} ) 
 \end{empheq}
where $\mathcal{X}\subset\R^{MN}$ is the closed and convex set consisting of points
satisfying the constraints in (\refeq{eq:c1}), (\refeq{eq:c2}), and (\refeq{eq:c3}). 
The problem in (\ref{eq:reducedProblem}) (which is a relaxation of problem (\refeq{eq:originalProbEP})-(\ref{eq:originalProb3EP}))
is the optimization problem we try to solve in this study. 
Unfortunately, it is still difficult to solve because we are
\textit{minimizing} a concave function. 
However, note that the concave objective function 
\begin{empheq}{align}
\label{eq:objective}
f(\ve{w}):=\sum_{i=1}^M\log(\epsilon+ \ve{s}_i^T\ve{w})
\end{empheq}
 is differentiable on $\mathcal{X}$ with gradient given by 
\begin{empheq}{align}
 \nabla f(\ve{w})=\sum_{i=1}^M \dfrac{\ve{s}_i}{\epsilon+\ve{s}_i^T
\ve{w}}. \nonumber
\end{empheq}
As a result, we can use the majorization-minimization (MM) algorithm (shown in 
appendix \ref{sec:Appendix}) to find a sequence of vectors $\ve{w}^{(n)}$ 
with non-increasing objective value; 
i.e., $f\left( \ve{w}^{(n+1)} \right) \leq f\left( \ve{w}^{(n)} \right)$. 
For sufficiently large $n$ we can expect to obtain solutions with good row 
sparsity patterns. 
More precisely, let
\begin{align}
  g(\ve{x},\ve{y}) = f(\ve{y})+\nabla f(\ve{y})^T(\ve{x}-\ve{y}), \nonumber
\end{align}
be the majorizing function of $f$ used by the MM algorithm. 
By doing so, the main iteration of the MM algorithm (which is obtained by substituting $g$ above into (\refeq{eq.mmit}) shown in the appendix) reduces to 
\begin{empheq}{align}
\label{eq:weightUpdate}
 \ve{w}^{(n+1)} \in \arg\min_{\ve{w}\in\mathcal{X}}\sum_{i=1}^M\dfrac{\ve{s}_i^T
\ve{w}}{\epsilon+\ve{s}_i^T\ve{w}^{(n)}}. 
\end{empheq}
where $\ve{w}^{(0)}$ is an arbitrary, feasible vector. The computation of the 
sequence from (\ref{eq:weightUpdate}) is the core task we perform in our algorithm
(summarized in Alg. \ref{alg:networkReconfiguration}).
Note that (\ref{eq:weightUpdate}) is a linear programming (LP) problem and hence it can 
be solved efficiently.

For the iteration in (\ref{eq:weightUpdate}), given a small arbitrary
value $\epsilon^\star>0$, we terminate the algorithm when either  
\begin{empheq}{align}
\label{eq:convergenceK}
  f \left(\ve{w}^{(n)} \right) - f \left( \ve{w}^{(n+1)} \right) < 
\epsilon^\star,
\end{empheq}
is valid or when the maximum number of iterations $\hat{n}$ is reached. Note that, 
even if we do not set the maximum number of iterations, the algorithm eventually 
terminates because $\left(f \left(\ve{w}^{(n)} \right)\right)$ converges (see the appendix). 
Upon termination, because of constraint (\ref{eq:c3}), we may obtain solutions 
$\ma{X}^\star$
where some entries are in the interval $\left(0,1\right)$
(though most entries of the results shown in section \ref{sec:numericalResults} 
are in $\{0,1\}$). Unfortunately, simply rounding those entries to the set
$\{0,1\}$ may lead to violations of the constraint in
(\ref{eq:c1}).
Hence, here we use the heuristic outlined in Alg. \ref{alg:userAllocation}. 
The main idea is to connect the users with entries from $\{0,1\}$ first. Then we 
try to connect the remaining users, starting from those corresponding to large entries $\omega_{i,j}$. New base stations are activated if the preceding operation fails.
\begin{algorithm}
  \caption{Network reconfiguration for improved energy efficiency}
\label{alg:networkReconfiguration}
\begin{algorithmic}[1]
  \REQUIRE set of all users, set of all base stations, constraints 
  \ENSURE optimized network configuration
\STATE initialize $\ve{w}^{(0)}$ with a feasible point
\REPEAT
\STATE compute $\ve{w}$ by solving the LP in (\ref{eq:weightUpdate})
\STATE increment $n$
\STATE update $\ve{w}^{(n)} := \ve{w}$
\UNTIL (\ref{eq:convergenceK}) is valid or $n=\hat{n}$
\STATE use a heuristic to map $\ve{w}^{(n)}$ from $\left[ 0,1 \right]^{M N}$
into $\{0,1\}^{M N}$.
\STATE connect the users to base stations according to $\ve{w}^{(n)}$.
\STATE deactivate all base stations no user is connected to.
\end{algorithmic}
\end{algorithm}

\begin{algorithm}
  \caption{User assignment to base stations}
\label{alg:userAllocation}
\begin{algorithmic}[1]
  \REQUIRE solution $\ma{X}^\star$, set of all users, set of all base stations, constraints 
  \ENSURE user assignment table
\STATE assign users with entries $x_{i,j}\in\{0,1\}$ to the corresponding base stations.
\STATE sort $x_{i,j}$ with values in $\left(0,1\right)$ in descending order
\FORALL{users $j$ in the sorted set}
\STATE assign user $j$ to base station $i$ corresponding to the largest entry 
$x_{i,j}$ that does not lead to a bandwidth violation at base station $i$.
\ENDFOR
\FORALL{users $j$ not assigned to any active base station }
\STATE activate the closest non-active base station and assign user $j$ to it.
\ENDFOR
\end{algorithmic}
\end{algorithm}

\textit{Remark:} The initialization of the MM algorithm is crucial to the overall 
performance. 
A bad initialization can lead to bad performance. 
We observed that a good starting point is to initialize the weights with a feasible 
solution $\ve{w}^{(0)}$ obtained by connecting each user to its closest base 
station. 
\section{Empirical evaluation}
\label{sec:numericalResults}
The simulation environment is similar to that in \cite{Niu2010CellZooming}, and it
consists of 100 cells 
in a 10 by 10 hexagonal cell layout. To avoid boundary effects, we use a 
wrap around model.
The inter cell distance is 500m, and all base stations have the same bandwidth limit 
$B = 5 \mbox{MHz}$. Furthermore, we use $c=400W$ as the typical power consumption for
a running base station.
For simplicity all users have the same rate requirement $r = 122 \mbox{kb/s}$.
New users are arriving according to a Poisson process with arrival rate $\lambda$. 
To get a spatially fluctuating user pattern, we define three hotspots with high user 
density. 
Those hotspots are normally distributed in the area, and they have a radius of $500$m.
New arriving users are dropped in a hotspot with a probability of 5\% each.  
Their location within the hotspot is normally distributed around the center. 
The remaining area has a uniform user distribution (including the hotspot areas).  
The spectral efficiency of each link from all base stations to all users
is calculated by (\ref{eq:specEffPL}), and the signal powers follow
the ITU propagation model for urban macro cell environments \cite{3GGPTR36814}. 
Unless otherwise stated, we use for the LP in (\ref{eq:weightUpdate}) $\epsilon = 10^{-3}$ and for 
the stopping criteria of the proposed 
algorithm $\hat{n}=20$ and $\epsilon^\star = 10^{-3}$.

\begin{figure}[ht]
	\centering
	\scalebox{0.58}{\includegraphics{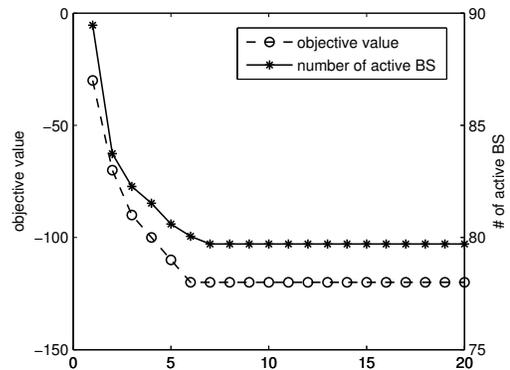}}
	\caption{Objective value and number of active base stations as a function of the number of iterations (for one realization). 
The  dashed curve shows the objective in (\ref{eq:objective}). 
The solid line shows the corresponding number of active base stations for the 
result of each iteration.}
	\label{fig:convergence}
\end{figure}

In Fig. \ref{fig:convergence} we show that the proposed algorithm typically
satisfies the stopping criterion (\ref{eq:convergenceK}) before reaching the
 maximum number of iterations.  Note that, for visual clarity, we let the 
algorithm run for at least $\hat{n}$ iterations in Fig. \ref{fig:convergence}, 
even if the alternative stopping criterion (\ref{eq:convergenceK}) is satisfied.
For a fixed user arrival rate leading to a mean number of 400 users in the 
network,  we have observed that, after six iterations, the objective value (\ref{eq:objective}) improves 
only marginally. 
In addition to the objective value, we also plot in Fig. \ref{fig:convergence} the cardinality of the set of 
active base stations obtained at each iteration. It can be seen that it follows the same trend.

Curves for other user arrival rates have shown a similar pattern. 
We have always observed a high decreasing rate of the objective value within approximately 
the first ten iterations and only a marginal decrease afterwards. 
More iterations have not improved the solution significantly in our scenario. 

\begin{figure}[ht]
	\centering
	\scalebox{0.6}{\includegraphics{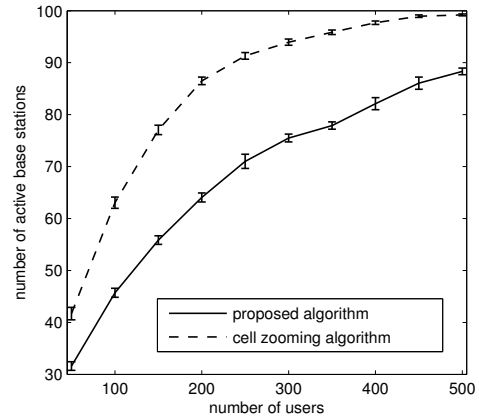}}
	\caption{Cardinality of active base station set for different user arrival  
rates $\lambda$ needed to provide QoS to all users. The results 
are averaged over ten realizations for each user arrival rate and the height of the error bars shows the estimated error of the mean.
The solid line shows our algorithm and the dashed line the cell zooming algorithm
from \cite{Niu2010CellZooming}}
	\label{fig:activeBS}
\end{figure}

In Fig. \ref{fig:activeBS} we compare the proposed algorithm with the centralized cell zooming approach in \cite{Niu2010CellZooming}.
The results are averaged over ten realizations for the same user arrival rate. 
The simulations show that the number of base stations increases as the user arrival
rate increases. 
It can be observed that, compared with the cell zooming algorithm, our algorithm gives 
solutions that satisfy the user constraints and use fewer base stations.
In addition, the proposed method has other major advantages over the cell zooming approach: 
\begin{itemize}
 \item It can be extended to heterogeneous networks, where we have different
power consumption for different network elements.
\item It can model the power consumption of the dynamic load, provided that it can be 
modeled as a concave or convex function.
\item It has a strong analytical justification for its good performance.
\end{itemize}

\section{Conclusions}
\label{sec:conclusion}
In this paper we have developed a novel approach to save energy in wireless networks 
by selecting a small number of active base stations that guarantees that the QoS 
(i.e., the minimum data rate) of all users is satisfied. We have shown that recent 
techniques that have been applied in, for example, compressed sensing can also be 
successfully applied in this application domain. In particular, the proposed algorithm needs 
to solve a simple LP at each iteration, so it can easily handle large-scale problems.
Simulations show that good sparse solutions are obtained with few iterations. 
We have also shown that our technique can outperform recent methods such as the 
cell-zooming approach \cite{Niu2010CellZooming} in practical scenarios. Finally, 
in stark contrast with the work in \cite{Niu2010CellZooming}, the proposed algorithm 
has other additional advantages such as a good analytical justification and the 
ability to consider heterogeneous networks and to consider the dynamic transmission power 
of base stations. 

\section*{\normalsize Acknowledgements}
This work has been partly supported by the framework of the research project ComGreen under the grant-number 01ME11010, 
which is funded by the German Federal Ministry of Economics and Technology (BMWi).
\appendix
\label{sec:Appendix}
\subsection{The majorization-minimization algorithm} 
For convenience, in this section we review the MM algorithm. 
The presentation here is based on \cite{sri11,hunter04}. \par
Suppose that we want to minimize a function $f:\mathcal{X}\rightarrow \R$, where 
$\mathcal{X}\subset\R^N$. 
In particular, in this study we assume that all optimization problems have a 
solution; i.e., there exists $\ve{x}^\star\in \mathcal{X}$ satisfying $\ve{x}^\star
\in\arg\min_{\ve{x}\in\mathcal{X}} f(\ve{x})$. 
Unless the optimization problem has a very special structure that can be exploited 
(e.g., convexity), finding such a point $\ve{x}^\star$ is computationally intractable 
in general, so we have to content ourselves with generating a sequence of vectors with 
non-increasing objective value, as explained below. 

A standard means of attaining small objective values is to apply the 
majorization-minimization (MM) technique \cite{hunter04}, which is a generalization of 
the celebrated expectation-maximization (EM) algorithm. 
In more detail, the MM algorithm is an iterative approach that tries to find a minimum 
of $f$ by minimizing at each iteration a surrogate function that i) majorizes 
$f$ at every point in $\mathcal{X}$ and that ii) is tangent to $f$ at the current 
estimate of a minimizer. 
More precisely, to apply the MM algorithm, we first need a function $g:\mathcal{X}
\times\mathcal{X}\rightarrow \R$ satisfying the following (see also \cite{sri11}):
\begin{align}
 \label{eq.mmp1}
 f(\ve{x})\le g(\ve{x},\ve{y}),\quad\forall\ve{x},\ve{y}\in
\mathcal{X}
\end{align}
\noindent and
\begin{align}
 \label{eq.mmp2}
 f(\ve{x})=g(\ve{x},\ve{x}),\quad\forall\ve{x}\in\mathcal{X}.
\end{align}
Then, starting from $\ve{x}^{(0)}\in\mathcal{X}$, the MM algorithm produces a sequence 
$\left(\ve{x}^{(n)}\right)\subset\mathcal{X}$ ($n\in\Natural$) by
\begin{align}
\label{eq.mmit}
 \ve{x}^{(n+1)} \in \arg\min_{\ve{x}\in\mathcal{X}}g(\ve{x},\ve{x}^{(n)}).
\end{align}

From the above, we see that the function $g$ should be sufficiently structured in order 
to make the optimization problem in (\refeq{eq.mmit}) easy to solve with efficient 
numerical approaches. 
In particular, if $f$ is concave and differentiable, a natural choice for $g$ is
\begin{align}
\label{eq.g_concave}
 g(\ve{x},\ve{y}) = f(\ve{y})+\nabla f(\ve{y})^T(\ve{x}-\ve{y}),
\end{align}
\noindent in which case the optimization problem in (\refeq{eq.mmit}) becomes a convex 
optimization problem provided that $\mathcal{X}$ is a convex set. 
This particular choice is common in, for example, sparse signal recovery \cite{candes08b}. 
\par 
More generally, irrespective of the choice of $f$ satisfying properties (\refeq{eq.mmp1}) 
and (\refeq{eq.mmp2}), we can easily verify that $\left(f(\ve{x}^{(n)})\right)$ is a monotone 
decreasing sequence:
\begin{multline*}
 f(\ve{x}^{(n+1)}) = g(\ve{x}^{(n+1)},\ve{x}^{(n+1)}) \\ \le g(\ve{x}^{(n+1)},
\ve{x}^{(n)}) \le g(\ve{x}^{(n)},\ve{x}^{(n)}) = f(\ve{x}^{(n)}),
\end{multline*}
where the equalities follow from (\refeq{eq.mmp2}), and the two inequalities follow from 
(\refeq{eq.mmp1}) and (\refeq{eq.mmit}), respectively. 
As a result, $f(\ve{x}^{(n)})\to c\in\R$ for some $c\ge f(\ve{x}^\star)$ as $n\to\infty$ 
(which in general does not imply the convergence of the sequence $\left(\ve{x}^{(n)}\right)$). 

\bibliographystyle{IEEEtran}
\bibliography{IEEEabrv,myreferences.bib}

\end{document}

%% file: macros.tex
\usepackage{dsfont}
\usepackage{amsmath}
\usepackage{amssymb}
\usepackage{empheq}
\usepackage{color}
\usepackage[pdftex]{graphicx}
\usepackage{algorithmic}
\usepackage[]{algorithm}

\graphicspath{{fig/}}

\newcommand{\field}[1]{\mathbb{#1}}
\newcommand{\ve}[1]{{\mathbf{#1}}}
\newcommand{\ma}[1]{{\mathbf{#1}}}
\newcommand{\set}[1]{{\mathcal{#1}}}

\newcommand{\R}{{\field{R}}}

\newcommand{\setm}{{\set{M}}}
\newcommand{\setn}{{\set{N}}}

\newcommand{\BEAS}{\begin{eqnarray*}}
\newcommand{\EEAS}{\end{eqnarray*}}
\newcommand{\BEQ}{\begin{equation}}
\newcommand{\EEQ}{\end{equation}}
\newcommand{\BIT}{\begin{itemize}}
\newcommand{\EIT}{\end{itemize}}
\newcommand{\BAR}{\begin{array}}
\newcommand{\EAR}{\end{array}}

\newcommand{\ones}{\mathbf 1}

\newcommand{\Natural}{{\mathbb N}}

\long\def\symbolfootnote[#1]#2{\begingroup%
 \def\thefootnote{\fnsymbol{footnote}}\footnote[#1]{#2}\endgroup} 

\newcounter{customlistcounter}
\newenvironment{customlist}[2]{
\setcounter{customlistcounter}{\value{#1}}
\begin{list}{#2}{\usecounter{#1}
\settowidth{\labelwidth}{#2}
\setlength{\leftmargin}{\labelwidth}
\addtolength{\leftmargin}{\labelsep}}
\setcounter{#1}{\value{customlistcounter}}
}{
\end{list}
}

\newcounter{assume}
\renewcommand{\theassume}{(A.\arabic{assume})}